# Pricing Financial Derivatives Subject to Counterparty Risk and Credit Value Adjustment

David Lee[1]

FinPricing

**ABSTRACT**

This article presents a generic model for pricing financial derivatives subject to counterparty credit risk. Both unilateral and bilateral types of credit risks are considered. Our study shows that credit risk should be modeled as American style options in most cases, which require a backward induction valuation. To correct a common mistake in the literature, we emphasize that the market value of a defaultable derivative is actually a risky value rather than a risk-free value. Credit value adjustment (CVA) is also elaborated. A practical framework is developed for pricing defaultable derivatives and calculating their CVAs at a portfolio level.

**Key Words**: credit value adjustment (CVA), credit risk modeling, risky valuation, collateralization, margin and netting.

---

[1] Email: david.lee@finpricing.com; Url: http://www.finpricing.com

A broad range of financial instruments bear credit risk. Credit risk may be unilateral or bilateral. Some derivatives such as, debt instruments (e.g., loans, bills, notes, bonds, etc), by nature contain only unilateral credit risk because only the default risk of one party appears to be relevant. Whereas some other derivatives, such as, over the counter (OTC) derivatives and securities financing transactions (SFT), bear bilateral credit risk because both parties are susceptible to default risk.

In the market, risk-free values are quoted for most financial derivatives. In other words, credit risk is not captured. Historical experience shows that credit risk often leads to significant losses. Therefore, it is obvious to all market participants that credit risk should be taken into account when reporting the fair value of any defaultable derivative. The adjustment to the risk-free value is known as the credit value adjustment (CVA). CVA offers an opportunity for banks to dynamically price credit risk into new trades and has become a common practice in the financial industry, especially for trading books. By definition, CVA is the difference between the risk-free value and the true (or risky or defaultable) value that takes into account the possibility of default. The risk-free value is what brokers quote or what trading systems or models normally report. The risky value, however, is a relatively less explored and less transparent area, which is the main challenge and core theme for credit risk measurement and management.

There are two primary types of models that attempt to describe default processes in the literature: structural models and reduced-form (or intensity) models. The structural models, studied by Merton (1974) and Longstaff and Schwartz (1995), regard default as an endogenous event, focusing on the capital structure of a firm. The reduced-form models introduced by Jarrow and Turnbull (1995) and Duffie and Singleton (1999) do not explain the event of default endogenously, but instead characterize it exogenously as a jump process. For pricing and hedging, reduced-form models are the preferred methodology.

Three different recovery models exist in the literature. The default payoff is either 1) a fraction of par (Madan and Unal (1998)), 2) a fraction of an equivalent default-free bond (Jarrow and Turnbull (1995)), or 3) a fraction of market value (Duffie and Singleton (1999)). The last one is most commonly used in the market. In their paper, Duffie and Singleton (1999) do not clearly state whether the market value of a defaultable derivative is a risky value or a risk-free value. Instead, the authors implicitly treat the market value as a risky value because the market price therein evolves in a defaultable manner. Otherwise, they cannot obtain the desired result. However, most of the later papers in the literature mistakenly think that the market value of a defaultable derivative is a risk-free value. Consequently, the results are incorrect. For instance, the current most popular CVA model described by Pykhtin and Zhu (2007), and Gregory (2009) is inappropriate in theory because the authors do not distinguish risky value and risk-free value when they conduct a risky valuation (i.e. valuing a defaultable derivative). In fact, the model has never been rigorously proved.

In this paper, we present generic models for valuing defaultable financial derivatives. For completeness, our study covers various cases: unilateral and bilateral, single payment and multiple payments, positive and negative payoffs. Although a couple of simple cases have been studied before by other authors, e.g. Duffie and Singleton (1999), Duffie and Huang (1996), who only provide heuristic derivations in a non-rigorous manner; analytic work on the other cases is novel. In contrast with the current recursive integral solution (see Duffie and Huang (1996)), our theory shows that the valuation of defaultable derivatives in most situations requires a backward induction procedure.

There is an intuitive way of understanding these backward induction behaviours: We can think that any defaultable derivative with bilateral credit risk embeds two default options. In other words, when entering a defaultable financial transaction, one party grants the other party an option to default and, at the same time, also receives an option to default itself. In theory, default may occur at any time. Therefore, default options are American style options that normally require a backward induction valuation.

We explicitly indicate that, within the context of risky valuation, the market value of a defaultable derivative is actually a risky value rather than a risk-free value, because a firm may default at any future time even if it survives today. An intuitive explanation is that after charging the upfront CVA, one already converted the market value of a defaultable derivative from the risk-free value to the risky value.

From a practical perspective, we propose a novel framework to perform the bilateral risky valuation and calculate the bilateral CVA at a portfolio level, which incorporates netting and margin (or collateral) agreements and captures wrong/right way risk.

The rest of this paper is organized as follows: Basic setup is provided in Section 1; pricing unilateral defaultable derivatives and their unilateral CVAs is discussed in Section 2; valuing bilateral asymmetric defaultable derivatives and their bilateral CVAs is elaborated on in Section 3; a practical framework that embraces netting agreements, margining agreements, and wrong/right way risk is proposed in Section 4, and the numerical results are presented in Section 5. The conclusions are given in Section 6.

## 1. Basic Setup

We consider a filtered probability space ($\Omega$, $\mathcal{F}$, $\{\mathcal{F}_t\}_{t\geq 0}$, $\mathcal{P}$) satisfying the usual conditions, where $\Omega$ denotes a sample space, $\mathcal{F}$ denotes a $\sigma$-algebra, $\mathcal{P}$ denotes a probability measure, and $\{\mathcal{F}_t\}_{t\geq 0}$ denotes a filtration.

The default model is based on the reduced-form approach proposed by Duffie and Singleton (1999) and Jarrow and Turnbell (1994), which does not explain the event of default endogenously, but characterizes it exogenously by a jump process. The stopping (or default) time $\tau$ is modeled as a Cox arrival process (also known as a doubly stochastic Poisson process) whose first jump occurs at default with a stochastic hazard rate or arrival intensity $h(t)$.

It is well-known that the survival probability, conditional on a realization path, from time $t$ to $s$ in this framework is defined by

$$S(t,s) := \Pr(\tau > s \mid \tau > t, Z) = \exp\left(-\int_t^s h(u)du\right) \tag{1a}$$

where $Z$ is the firm-specific state process that helps forecast that firm's default. The default probability, conditional on the realization path, for the period ($t$, $s$) in this framework is defined by

$$Q(t,s) = \Pr(\tau \leq s \mid \tau > t, Z) = 1 - S(t,s) = 1 - \exp\left(-\int_t^s h(u)du\right) \tag{1b}$$

Applying the law of iterated expectations, we express the expected survival probability for the period $(t, s)$ as

$$\bar{S}(t,s) = E\left[\exp\left(-\int_t^s h(u)du\right)\bigg|\mathcal{F}_t\right] \tag{2a}$$

where $E\{\bullet|\mathcal{F}_t\}$ is the expectation conditional on the $\mathcal{F}_t$. The expected default probability for the period $(t, s)$ is expressed as

$$\bar{Q}(t,s) = 1 - \bar{S}(t,s) = 1 - E\left[\exp\left(-\int_t^s h(u)du\right)\bigg|\mathcal{F}_t\right] \tag{2b}$$

There are three different recovery models in the literature. The default payoff is either 1) a fraction of par, 2) a fraction of an equivalent default-free bond, or 3) a fraction of market value. We use the recovery model of market value in our study.

The binomial default rule considers only two possible states: default or survival. For a discrete one-period $(t, s)$ economy, at the end of the period a defaultable derivative either defaults with the default probability $Q(t,s)$ or survives with the survival probability $S(t,s)$. Assume that the market value of the defaultable derivative at time $s$ is $V^D(s)$. The default payoff is a fraction of the market value given by $\varphi(s)V^D(s)$, where $\varphi(s)$ is the default recovery rate at time $s$, whereas the survival payoff is equal to the market value $V^D(s)$ itself. Therefore, the risky value of the derivative, conditional on the realization path, is the discounted expectation of the payoffs, given by

$$V^D(t) = D(t,s)\big(Q(t,s)\varphi(s)V^D(s) + S(t,s)V^D(s)\big) \tag{3a}$$

where

$$D(t,s) = \exp\left[-\int_t^s r(u)du\right] \tag{3b}$$

where $D(t,s)$ denotes the stochastic risk-free discount factor at $t$ for the maturity $s$ and $r(u)$ denotes the risk-free short rate at time $u$ ($t \leq u \leq s$).

If the derivative survives at time *s*, the survival value is equal to the market value. However, the derivative may default at any time after *s*. Therefore, the present survival value or the market value at time *s* is a risky value. In other words, for risk-free valuations, the market value is a risk-free value, but for risky valuations, the market value is actually a risky value. An intuitive explanation is that one already converted the market value from the risk-free value to the risky value after charging the upfront CVA.

## 2. Unilateral Defaultable Derivatives Valuation and Unilateral CVA

Unilateral risky valuation can be applied to two cases. By nature, the derivatives in case 1 involve only unilateral credit risk. In case 2, the derivatives actually bear bilateral credit risk, but people sometimes may treat bilateral defaultable derivatives as unilateral ones for simplicity.

Two parties are denoted as *A* and *B*. The unilateral credit risk assumes that only one party is defaultable and the other one is default-free. In this section, we assume that party *A* is default-free, whereas party *B* is defaultable. All calculations are from the perspective of party *A*. Let the valuation date be *t*. We study single payment cases first.

### 2.1. Single Payment Cases

Consider a defaultable derivative that promises to pay a $X_T$ from party *B* to party *A* at maturity date *T*, and nothing before date *T*. The payoff $X_T$ may be positive or negative, i.e. the derivative may be either an asset or a liability to each party. The risk-free value of the derivative at valuation date *t* is given by

$$V^F(t) = E\{D(t,T)X_T | \mathcal{F}_t\} \tag{4}$$

where $D(t,T)$ is the risk-free discount factor defined in (3b).

We divide the time period (*t, T*) into *n* very small time intervals ($\Delta t$) and assume that a default may occur only at the end of each very small period.

**Proposition 1:** *The unilateral risky value of the single payment derivative is given by*

$$V^D(t) = E\{C_B(t,T)X_T | \mathcal{F}_t\} \qquad (5a)$$

*where*

$$C_B(t,T) = \exp\left[-\sum_{i=0}^{n-1} c_B(t+i\Delta t)\Delta t\right] \qquad (5b)$$

$$c_B(t+i\Delta t) = r(t+i\Delta t) + 1_{V^D(t+(i+1)\Delta t) \geq 0} h_B(t+i\Delta t)(1-\varphi_B(t+i\Delta t)) \qquad (5c)$$

*where $1_Y$ is an indicator function that is equal to one if Y is true and zero otherwise.*

Proof: See the Appendix.

We may think of $C_B(t,T)$ as the risk-adjusted discount factor and $c_B(u)$ as the risk-adjusted short rate. Here $\varphi_B(u)$ represents the default recovery rate of party B, and $h_B(u)$ represents the hazard rate of party B. $s_B(u) = h_B(u)(1-\varphi_B(u))$ is called the credit spread or the short credit spread.

Compared to the risk-free valuation (4), the risky valuation (5) is substantially complex. The intermediate values are vital to determine the final price. For a small time interval, the current risky value has a dependence on the future risky value. Only on the final payment date *T*, the value of the derivative and the maximum amount of information needed to determine the risk-adjusted discount factor are revealed. This type of problem can be best solved by working backwards in time, with the later risky value feeding into the earlier ones, so that the process builds on itself in a recursive fashion, which is referred to as backward induction. The most popular backward induction valuation algorithms are lattice/tree and regression-based Monte Carlo.

For an intuitive explanation, we can posit that a defaultable derivative under the unilateral credit risk assumption has an embedded default option. In other words, one party entering a defaultable financial transaction actually grants the other party an option to default. If we assume that a default may occur at any time, the default option is an American style option. American options normally have backward recursive natures and require backward induction valuations.

In theory, a default may happen at any time. The Continuous-Time Risky Valuation Model, or simply the Continuous-Time Model (CTM), assumes that a defaultable derivative is continuously defaultable. Proposition 1 can be further expressed in the form of the CTM as:

***Corollary 1.1***: *The unilateral risky value of the single payment derivative under the CTM is given by*

$$V^D(t) = E\{\overline{C}_B(t,T)X_T | \mathcal{F}_t\} \qquad (6a)$$

*where*

$$\overline{C}_B(t,T) = \exp\left[-\int_t^T \overline{c}_B(u)du\right] \qquad (6b)$$

$$\overline{c}_B(u) = r(u) + 1_{V^D(u) \geq 0} h_B(u)(1 - \varphi_B(u)) \qquad (6c)$$

The proof of this corollary becomes straightforward, according to Proposition 1, by taking the limit as $\Delta t$ approaches zero.

Proposition 1 provides a general form for pricing a unilateral defaultable single payment derivative. Applying it to a particular situation in which we assume that the payoff $X_T$ is non-negative, we derive the following corollary:

***Corollary 1.2***: *If the payoff $X_T$ is non-negative, the risky value of the single payment derivative under the CTM is given by*

$$V^D(t) = E\{\breve{C}_B(t,T)X_T | \mathcal{F}_t\} \qquad (7a)$$

*where*

$$\breve{C}_B(t,T) = \exp\left[-\int_t^T \breve{c}_B(u)du\right] \qquad (7b)$$

$$\breve{c}_B(u) = r(u) + h_B(u)(1 - \varphi_B(u)) \qquad (7c)$$

The proof of this corollary is easily obtained according to (6) by setting $V^D(u) \geq 0$. Since the payoff $X_T$ is non-negative, the intermediate value $V^D(u)$ is also non-negative.

Equation (7) is the same as Equation (10) in Duffie and Singleton (1999). In their heuristic derivation, the authors implicitly treat the market value of a defaultable derivative as a risky value because the market price therein evolves in a defaultable manner. Otherwise, they would not have been able to obtain the desired result.

The risky valuation becomes quite simple when the payoff is non-negative. This corollary says that a single non-negative payment defaultable derivative can be priced using the present value of the promised payoff $X_T$ discounted by the risk-adjusted discount rate $\breve{c}_B(u)$ instead of the risk-free rate $r(u)$.

By definition, the corresponding CVA of the single non-negative payment derivative can be expressed as

$$CVA^U(t) = V^F(t) - V^D(t) = E\{D(t,T)X_T|\mathcal{F}_t\} - E\{\breve{C}_B(t,T)X_T|\mathcal{F}_t\} \tag{8}$$

Since $\breve{C}_B(t,T)$ is always smaller than $D(t,T)$, the CVA is, in this case, a credit charge. The credit charge covering party *A*'s potential loss comes from the scenario of party *B*'s default when party *A* is in the money in the portfolio.

Under the CTM, a unilateral defaultable derivative has an embedded American style default option, which is continuously defaultable. Since American options are difficult to hedge, a common practice in the market is to use Bermudan options to approximate American ones. Consequently, we assume that a default may only happen at some discrete times. A natural selection is to assume that a default may occur only on the payment dates. The Discrete-Time Risky Valuation Model, or simply the Discrete-Time Model (DTM), assumes that a defaultable derivative may only default on payment dates.

**Proposition 2:** *The unilateral risky value of the single payment derivative under the DTM is given by*

$$V^D(t) = E[G_B(t,T)X_T|\mathcal{F}_t] \tag{9a}$$

*where*

$$G_B(t,T) = D(t,T)[1 - 1_{X_T \geq 0}Q_B(t,T)(1-\varphi_B(T))] \tag{9b}$$

Proof: See the Appendix.

Here we may consider $G_B(t,T)$ as the risk-adjusted discount factor. Proposition 2 states that the unilateral risky valuation of the single payoff derivative has a dependence on the sign of the payoff. If the payoff is positive, the risky value is equal to the risk-free value minus the discounted potential loss. Otherwise, the risky value is equal to the risk-free value.

The corresponding unilateral CVA of the single payment derivative under the DTM can be expressed as

$$CVA^U(t) = V^F(t) - V^D(t) = E\{D(t,T)X_T | \mathcal{F}_t\} - E\{G_B(t,T)X_T | \mathcal{F}_t\}$$
$$= E[1_{X_T \geq 0} D(t,T)(1 - \varphi_B(T))Q_B(t,T)X_T | \mathcal{F}_t] \qquad (10)$$

Equation (10) shows that if the payoff is in the money, the CVA is a credit charge equal to the discounted potential loss. If the payoff is out of the money, the CVA is zero.

Proposition 2 has a general form that applies in a particular situation in which we assume that the payoff $X_T$ is non-negative.

***Corollary 2.1:*** *If the payoff $X_T$ is non-negative, the risky value of the single payment derivative under the DTM is given by*

$$V^D(t) = E[\overline{G}_B(t,T)X_T | \mathcal{F}_t] \qquad (11a)$$

*where*

$$\overline{G}_B(t,T) = D(t,T)[1 - Q_B(t,T)(1 - \varphi_B(T))] = D(t,T)(S_B(t,T) + \varphi_B(T)Q_B(t,T)) \qquad (11b)$$

The proof of this corollary is straightforward according to Proposition 2 by setting $X_T \geq 0$. Equations (11) are consistent with equations (3).

### 2.2. Multiple Payments Cases

Suppose that a defaultable derivative has *m* cash flows. Any of these cash flows may be positive or negative. Let the *m* cash flows be represented as $X_1, \ldots, X_m$ with payment dates $T_1, \ldots, T_m$. The risk-free price of the derivative is given by

$$V^F(t) = \sum_{i=1}^{m} E\{D(t,T_i)X_i|\mathcal{F}_t\} \tag{12}$$

We divide any payment date period ($T_{i-1}$, $T_i$) into $n_i$ very small time intervals ($\Delta t$) and assume that a default may occur only at the end of each very small period.

**Proposition 3:** *The unilateral risky value of the multiple payments derivative is given by*

$$V^D(t) = \sum_{i=1}^{m} E\left[(C_B(t,T_i))X_i|\mathcal{F}_t\right] \tag{13a}$$

*where*

$$C_B(t,T_i) = \exp\left[-\sum_{j=0}^{\sum_{k=1}^{i}n_k - 1} c_B(t+j\Delta t)\Delta t\right] \tag{13b}$$

$$c_B(t+j\Delta t) = r(t+j\Delta t) + 1_{V^D(t+(j+1)\Delta t)\geq 0} h_B(t+j\Delta t)(1 - \varphi_B(t+j\Delta t)) \tag{13c}$$

Proof: See the Appendix.

Proposition 3 says that the pricing process of a multiple payments derivative has a backward nature since there is no way of knowing which risk-adjusted discounting rate should be used without knowledge of the future value. In other words, the present value takes into account the results of all future decisions. On the final payment date, the value of a derivative and the decision strategy are clear. Therefore, the evaluation must be done in a backward fashion, working from the final payment date towards the present. This type of valuation process is referred to as backward induction.

If we model default in a continuous-time setting, the Proposition 3 can be further expressed as follows.

***Corollary 3.1***: *The unilateral risky value of the multiple payments derivative under the CTM is given by*

$$V^D(t) = \sum_{i=1}^{m} E\left[(\overline{C}_B(t,T_i))X_i|\mathcal{F}_t\right] \tag{14}$$

*where*

$$\overline{C}_B(t,T_i) = \exp\left[-\int_t^{T_i} \overline{c}_B(u)du\right] \tag{14b}$$

$$\overline{c}_B(u) = r(u) + 1_{V^D(u)\geq 0} h_B(u)(1 - \varphi_B(u)) \tag{14c}$$

The proof of this corollary is easily obtained, according to Proposition 3, by taking the limit as $\Delta t$ approaches zero.

Proposition 3 has a general form that applies in a particular situation in which we assume that all the payoffs are non-negative.

***Corollary 3.2***: *If all the payoffs are non-negative, the risky value of the multiple payments derivative under the CTM is given by*

$$V^D(t) = \sum_{i=1}^{m} E\left[\left(\breve{C}_B(t,T_i)\right)X_i \big| \mathcal{F}_t\right] \qquad (15a)$$

*where*

$$\breve{C}_B(t,T_i) = \exp\left[-\int_t^{T_i} \breve{c}_B(u)du\right] \qquad (15b)$$

$$\breve{c}_B(u) = r(u) + h_B(u)(1 - \varphi_B(u)) \qquad (15c)$$

The proof of this corollary is straightforward, according to (14), by setting $V^D(u) \geq 0$.

This is the formula for pricing defaultable bonds in the market. Corollary 3.2 says that if all the payoffs are positive, we can evaluate each payoff separately and sum the corresponding results. In other words, payoffs in this case can be treated as independent.

If we assume that a default may occur only on the payment dates, the result is the following proposition in a discrete-time setting.

**Proposition 4:** T*he unilateral risky value of the multiple payments derivative under the DTM is given by*

$$V^D(t) = \sum_{i=1}^{m} E\left[\prod_{j=0}^{i-1} G_B(T_j,T_{j+1})X_i \big| \mathcal{F}_t\right] \qquad (16a)$$

where $t = T_0$ and

$$G_B(T_j,T_{j+1}) = D(T_j,T_{j+1})\left[1 - 1_{(X_{j+1}+V^D(T_{j+1}))\geq 0} Q_B(T_j,T_{j+1})(1-\varphi_B(T_{j+1}))\right] \qquad (16b)$$

Proof: See the Appendix.

Similar to Proposition 3, the individual payoffs under Proposition 4 cannot be evaluated separately. The current risky value depends on the future risky value. This type of problem is usually solved using backward induction algorithms.

Proposition 4 has a general form that applies in a particular situation where we assume that all payoffs are non-negative.

***Corollary 4.1***: *If all the payoffs are non-negative, the risky value of the multiple payments derivative under the DTM is given by*

$$V^D(t) = \sum_{i=1}^{m} E\left[\left(\prod_{j=0}^{i-1} \overline{G}_B(T_j, T_{j+1})\right) X_i \big| \mathcal{F}_t\right] \quad (17a)$$

Where $t = T_0$ and

$$\overline{G}_B(T_j, T_{j+1}) = D(T_j, T_{j+1})\left[1 - Q_B(T_j, T_{j+1})(1 - \varphi_B(T_{j+1}))\right] \quad (17b)$$

The proof of this corollary is easily obtained according to Proposition 4 by setting $\left(X_{j+1} + V^D(T_{j+1})\right) \geq 0$.

## 3. Bilateral Defaultable Derivative Valuation and Bilateral CVA

A critical ingredient of the pricing of a bilateral defaultable derivative is the rules for settlement in default. There are two rules in the market. The *one-way payment rule* was specified by the early International Swap Dealers Association (ISDA) master agreement. The non-defaulting party is not obligated to compensate the defaulting party if the remaining market value of the derivative is positive for the defaulting party. The *two-way payment rule* is based on current ISDA documentation. In the event of default, if the contract has positive value to the non-defaulting party, the defaulting party pays a fraction of the market value of the derivative to the non-defaulting party. If the contract has positive value to the defaulting party, the non-defaulting party will pay the full market value of the derivative to the defaulting party. Within the context of risky valuation, one should consider the market value of a defaultable derivative as a risky value.

The default indictor $\overset{\text{)}}{j}$ for party $j$ ($j=A, B$) is a random variable with a Bernoulli distribution, which takes value 1 with default probability $Q_j$, and value 0 with survival probability $S_j$. Consider a pair of random variables ($\overset{\text{)}}{A}$, $\overset{\text{)}}{B}$) that has a bivariate Bernoulli distribution as summarized in Table 1.

**Table 1. Bivariate Bernoulli Distribution**

This table shows the joint and marginal distributions of a bivariate Bernoulli distribution of $\overset{\text{)}}{A}$ and $\overset{\text{)}}{B}$. Marginally, each random variable $\overset{\text{)}}{j}$ ($\overset{\text{)}}{j} = \overset{\text{)}}{A}, \overset{\text{)}}{B}$) follows a univariate Bernoulli distribution that takes value 1 with default probability $Q_j$ and value 0 with survival probability $S_j$. $\gamma = \rho\sqrt{Q_A S_A Q_B S_B}$ where $\rho$ is the correlation coefficient of $\overset{\text{)}}{A}$ and $\overset{\text{)}}{B}$.

|  |  | Joint Distribution | | Marginal Distribution |
|---|---|---|---|---|
|  |  | $\overset{\text{)}}{A}=1$ | $\overset{\text{)}}{A}=0$ |  |
| **Joint Distribution** | $\overset{\text{)}}{B}=1$ | $Q_B Q_A + \gamma$ | $Q_B S_A - \gamma$ | $Q_B$ |
|  | $\overset{\text{)}}{B}=0$ | $S_B Q_A - \gamma$ | $S_B S_A + \gamma$ | $S_B$ |
| **Marginal Distribution** |  | $Q_A$ | $S_A$ |  |

### 3.1. Single Payment Cases

Consider a defaultable derivative that promises to pay a $X_T$ from party B to party A at maturity date $T$ and nothing before date $T$. The payoff may be either an asset or a liability to each party.

We divide the time period ($t, T$) into $n$ very small time intervals ($\Delta t$) and assume that a default may occur only at the end of each very small period.

**Proposition 5:** *The bilateral risky value of the single payment derivative is given by*

$$V^D(t) = E\{O(t,T)X_T | \mathcal{F}_t\} \tag{18a}$$

*where*

$$O(t,T) = \exp\left(-\sum_{i=0}^{n-1} o(t+i\Delta t)\Delta t\right) \qquad (18b)$$

$$o(t+i\Delta t) = r(t+i\Delta t) + 1_{V(t+(i+1)\Delta t)\geq 0} p_B(t+i\Delta t) + 1_{V(t+(i+1)\Delta t)<0} p_A(t+i\Delta t) \qquad (18c)$$

$$\begin{aligned}
p_B(t+i\Delta t) &= (1-\varphi_B(t+i\Delta t))h_B(t+i\Delta t) + (1-\overline{\varphi}_B(t+i\Delta t))h_A(t+i\Delta t) \\
&\quad - (1-\varphi_B(t+i\Delta t) - \overline{\varphi}_B(t+i\Delta t) + \varphi_{AB}(t+i\Delta t)) \\
&\quad \times \left[\rho\sqrt{h_A(t+i\Delta t)h_B(t+i\Delta t)(1-h_B(t+i\Delta t)\Delta t)(1-h_A(t+i\Delta t)\Delta t)} \right. \\
&\quad \left. + h_B(t+i\Delta t)h_A(t+i\Delta t)\Delta t\right]
\end{aligned} \qquad (18d)$$

$$\begin{aligned}
p_A(t+i\Delta t) &= (1-\varphi_A(t+i\Delta t))h_A(t+i\Delta t) + (1-\overline{\varphi}_A(t+i\Delta t))h_B(t+i\Delta t) \\
&\quad - (1-\varphi_A(t+i\Delta t) - \overline{\varphi}_A(t+i\Delta t) + \varphi_{AB}(t+i\Delta t)) \\
&\quad \times \left[\rho\sqrt{h_A(t+i\Delta t)h_B(t+i\Delta t)(1-h_B(t+i\Delta t)\Delta t)(1-h_A(t+i\Delta t)\Delta t)} \right. \\
&\quad \left. + h_B(t+i\Delta t)h_A(t+i\Delta t)\Delta t\right]
\end{aligned} \qquad (18e)$$

Proof: See the Appendix.

We may think of $O(t,T)$ as the bilateral risk-adjusted discount factor and $o(u)$ as the bilateral risk-adjusted short rate; $\varphi_j$ ($j=A, B$) represents the default recovery rate of party $j$, i.e. the fraction of the market value paid by the defaulting party $j$ when the market value is negative for $j$; $h_j$ represents the hazard rate of party j; $\overline{\varphi}_j$ represents the non-default recovery rate of party $j$, i.e. the fraction of the market value paid by non-defaulting party $j$ when the market value is negative for $j$. $\overline{\varphi}_j = 0$ represents the one-way settlement rule, while $\overline{\varphi}_j = 1$ represents the two-way settlement rule. $\rho$ denotes the default correlation coefficient of A and B. $\varphi_{AB}$ denotes the joint recovery rate when both parties A and B default simultaneously.

For any time interval ($u$, $u+\Delta t$), the bilateral risk-adjusted short rate $o(u)$ has a switching-type dependence on the sign of future value $V^D(u+\Delta t)$. Similar to Proposition 1, the valuation process given by Proposition 5 builds on itself in a backward recursive fashion and requires a backward induction valuation.

Proposition 5 provides a general form for pricing a bilateral risky single payment derivative. Applying it to a particular situation in which we assume that parties A and B do not default simultaneously and have independent default risks, i.e. $\rho = 0$ and $\varphi_{AB} = 0$, we derive the following corollary.

**Corollary 5.1:** *If parties A and B do not default simultaneously and have independent default risks, the bilateral risky value of the single payment derivative under the CTM is given by*

$$V^D(t) = E\{\overline{O}(t,T)X_T | \mathcal{F}_t\} \tag{19a}$$

*where*

$$\overline{O}(t,T) = \exp\left[-\int_t^T \overline{o}(u)du\right] \tag{19b}$$

$$\overline{o}(u) = r(u) + 1_{V^D(u) \geq 0}\overline{p}_B(u) + 1_{V^D(u) < 0}\overline{p}_A(u) \tag{19c}$$

$$\overline{p}_B(u) = (1-\varphi_B(u))h_B(u) + (1-\overline{\varphi}_B(u))h_A(u) \tag{19d}$$

$$\overline{p}_A(u) = (1-\varphi_A(u))h_A(u) + (1-\overline{\varphi}_A(u))h_B(u) \tag{19e}$$

The proof of this corollary becomes straightforward according to Proposition 5 by setting $\rho = 0$ and $\varphi_{AB} = 0$, taking the limit as $\Delta t$ approaches zero, and having $h_B(u)h_A(u)\Delta t^2 \approx 0$ for very small $\Delta t$.

Corollary 5.1 is the same as equation (2.5') in Duffie and Huang (1996), but their derivation is heuristic rather than rigorous.

The corresponding bilateral CVA of the single payment derivative in this case can be expressed as

$$CVA^B(t) = V^F(t) - V^D(t) = E\{D(t,T)X_T | \mathcal{F}_t\} - E\{\overline{O}(t,T)X_T | \mathcal{F}_t\} \tag{20}$$

Since $\overline{O}(t,T)$ is always smaller than $D(t,T)$, the bilateral CVA may be positive or negative depending on the sign of the payoff. In other words, the CVA may be either a credit charge or a credit benefit. A credit charge that is the value of a default loss comes from the scenario of one's counterparty defaulting when one is in the money, while a credit benefit is the value of a default gain in the scenario of one's own default when one is out of money.

If we assume that a default may occur only on the payment dates, the default options become either European options or Bermudan options.

**Proposition 6:** *The bilateral risky value of the single payment derivative under the DTM is given by*

$$V^D(t) = E[Y(t,T)X_T | \mathcal{F}_t] \quad (21a)$$

*where*

$$Y(t,T) = D(t,T)\left(1_{X_T \geq 0} y_B(t,T) + 1_{X_T < 0} y_A(t,T)\right) \quad (21b)$$

$$\begin{aligned} y_B(t,T) &= S_B(t,T)S_A(t,T) + \varphi_B(T)Q_B(t,T)S_A(t,T) + \overline{\varphi}_B(T)S_B(t,T)Q_A(t,T) \\ &+ \varphi_{AB}(T)Q_B(t,T)Q_A(t,T) + \gamma(t,T)\left(1 - \varphi_B(T) - \overline{\varphi}_B(T) + \varphi_{AB}(T)\right) \end{aligned} \quad (21c)$$

$$\begin{aligned} y_A(t,T) &= S_B(t,T)S_A(t,T) + \varphi_A(T)Q_A(t,T)S_B(t,T) + \overline{\varphi}_A(T)S_A(t,T)Q_B(t,T) \\ &+ \varphi_{AB}(T)Q_B(t,T)Q_A(t,T) + \gamma(t,T)\left(1 - \varphi_B(T) - \overline{\varphi}_B(T) + \varphi_{AB}(T)\right) \end{aligned} \quad (21d)$$

$$\gamma(t,T) = \rho\sqrt{S_B(t,T)Q_B(t,T)S_A(t,T)Q_A(t,T)} \quad (21e)$$

Proof: See the Appendix.

We may think of $Y(t,T)$ as the risk-adjusted discount factor. Proposition 6 tells us that the bilateral risky price of a single payment derivative can be expressed as the present value of the payoff discounted by a risk-adjusted discount factor that has a switching-type dependence on the sign of the payoff.

Proposition 6 has a general form that applies in a particular situation where we assume that parties *A* and *B* do not default simultaneously and have independent default risks, i.e. $\rho = 0$ and $\varphi_{AB} = 0$.

**Corollary 6.1:** *If parties A and B do not default simultaneously and have independent default risks, the bilateral risky value of the single payment derivative under the DTM is given by*

$$V^D(t) = E\{\overline{Y}(t,T)X_T | \mathcal{F}_t\} \quad (22a)$$

*where*

$$\bar{Y}(t,T) = D(t,T)\left(1_{X_T \geq 0} \bar{y}_B(t,T) + 1_{X_T < 0} \bar{y}_A(t,T)\right) \quad (22b)$$

$$\bar{y}_B(t,T) = S_B(t,T)S_A(t,T) + \varphi_B(T)Q_B(t,T)S_A(t,T) + \bar{\varphi}_B(T)S_B(t,T)Q_A(t,T) \quad (22c)$$

$$\bar{y}_A(t,T) = S_B(t,T)S_A(t,T) + \varphi_A(T)Q_A(t,T)S_B(t,T) + \bar{\varphi}_A(T)S_A(t,T)Q_B(t,T) \quad (22d)$$

The proof of this corollary is easily obtained according to Proposition 6 by setting $\rho = 0$ and $\varphi_{AB} = 0$.

### 3.2. Multiple Payments Cases

Suppose that a defaultable derivative has $m$ cash flows. Let the $m$ cash flows be represented as $X_1, \ldots, X_m$ with payment dates $T_1, \ldots, T_m$. Each cash flow may be positive or negative.

We divide any payment date period ($T_{i-1}$, $T_i$) into $n_i$ very small time intervals ($\Delta t$) and assume that a default may occur only at the end of each very small period.

**Proposition 7:** *The bilateral risky value of the multiple payments derivative is given by*

$$V^D(t) = \sum_{i=1}^{m} E\left[(O(t,T_i))X_i \big| \mathcal{F}_t\right] \quad (23a)$$

*where*

$$O(t,T_i) = \exp\left[-\sum_{j=0}^{\sum_{k=1}^{i} n_k - 1} o(t + j\Delta t)\Delta t\right] \quad (23b)$$

$$o(t + j\Delta t) = r(t + j\Delta t) + 1_{V(t+(j+1)\Delta t) \geq 0} p_B(t + j\Delta t) + 1_{V(t+(j+1)\Delta t) < 0} p_A(t + j\Delta t) \quad (23c)$$

$$\begin{aligned} p_B(t + j\Delta t) &= (1 - \varphi_B(t + j\Delta t))h_B(t + j\Delta t) + (1 - \bar{\varphi}_B(t + j\Delta t))h_A(t + j\Delta t) \\ &\quad - (1 - \varphi_B(t + j\Delta t) - \bar{\varphi}_B(t + j\Delta t) + \varphi_{AB}(t + j\Delta t)) \\ &\quad \times \left[\rho\sqrt{h_A(t + j\Delta t)h_B(t + j\Delta t)(1 - h_B(t + j\Delta t)\Delta t)(1 - h_A(t + j\Delta t)\Delta t)} \right. \\ &\quad \left. + h_B(t + j\Delta t)h_A(t + j\Delta t)\Delta t\right] \end{aligned} \quad (23d)$$

$$\begin{aligned} p_A(t + j\Delta t) &= (1 - \varphi_A(t + j\Delta t))h_A(t + j\Delta t) + (1 - \bar{\varphi}_A(t + j\Delta t))h_B(t + j\Delta t) \\ &\quad - (1 - \varphi_A(t + j\Delta t) - \bar{\varphi}_A(t + j\Delta t) + \varphi_{AB}(t + j\Delta t)) \\ &\quad \times \left[\rho\sqrt{h_A(t + j\Delta t)h_B(t + j\Delta t)(1 - h_B(t + j\Delta t)\Delta t)(1 - h_A(t + j\Delta t)\Delta t)} \right. \\ &\quad \left. + h_B(t + j\Delta t)h_A(t + j\Delta t)\Delta t\right] \end{aligned} \quad (23e)$$

Proof: See the Appendix.

Similar to Proposition 3, the individual payoffs under Proposition 7 are coupled and cannot be evaluated separately. The process requires a backward induction valuation.

Proposition 7 provides a general form for pricing a bilateral multiple payment derivative. Applying it to a particular situation in which we assume that parties *A* and *B* do not default simultaneously and have independent default risks, i.e. $\rho = 0$ and $\varphi_{AB} = 0$, we derive the following corollary.

**Corollary 7.1:** *If parties A and B do not default simultaneously and have independent default risks, the bilateral risky value of the single payment derivative under the CTM is given by*

$$V^D(t) = \sum_{i=1}^{m} E\left[\left(\overline{O}(t,T_i)\right) X_i \big| \mathcal{F}_t\right] \tag{24}$$

*where $\overline{O}(t,T_i)$ is defined in (19).*

The proof of this corollary becomes straightforward according to Proposition 7 by setting $\rho = 0$ and $\varphi_{AB} = 0$, taking the limit as $\Delta t$ approaches zero, and having $h_B(u)h_A(u)\Delta t^2 \approx 0$ for very small $\Delta t$.

The default options of a defaultable derivative under the CTM are American style options, while the default options under the DTM are Bermudan style ones.

**Proposition 8:** *The bilateral risky value of the multiple payments derivative under the DTM is given by*

$$V^D(t) = \sum_{i=1}^{m} E\left[\left(\prod_{j=0}^{i-1} Y(T_j, T_{j+1})\right) X_i \big| \mathcal{F}_t\right] \tag{25a}$$

*where $t = T_0$ and*

$$Y(T_j, T_{j+1}) = D(T_j, T_{j+1})\left(1_{(X_{j+1}+V^D(T_{j+1}))\geq 0} y_B(T_j, T_{j+1}) + 1_{(X_{j+1}+V^D(T_{j+1}))<0} y_A(T_j, T_{j+1})\right) \tag{25b}$$

$$\begin{aligned} y_B(T_j, T_{j+1}) = &S_B(T_j, T_{j+1})S_A(T_j, T_{j+1}) + \varphi_B(T_{j+1})Q_B(T_j, T_{j+1})S_A(T_j, T_{j+1}) + \overline{\varphi}_B(T_{j+1})S_B(T_j, T_{j+1})Q_A(T_j, T_{j+1}) \\ &+ \varphi_{AB}(T_{j+1})Q_B(T_j, T_{j+1})Q_A(T_j, T_{j+1}) + \gamma(T_j, T_{j+1})\left(1 - \varphi_B(T_{j+1}) - \overline{\varphi}_B(T_{j+1}) + \varphi_{AB}(T_{j+1})\right) \end{aligned} \tag{25c}$$

$$\begin{aligned} y_A(T_j, T_{j+1}) = &S_B(T_j, T_{j+1})S_A(T_j, T_{j+1}) + \varphi_A(T_{j+1})Q_A(T_j, T_{j+1})S_B(T_j, T_{j+1}) + \overline{\varphi}_A(T_{j+1})S_A(T_j, T_{j+1})Q_B(T_j, T_{j+1}) \\ &+ \varphi_{AB}(T_{j+1})Q_B(T_j, T_{j+1})Q_A(T_j, T_{j+1}) + \gamma(T_j, T_{j+1})\left(1 - \varphi_B(T_{j+1}) - \overline{\varphi}_B(T_{j+1}) + \varphi_{AB}(T_{j+1})\right) \end{aligned} \tag{25d}$$

$$\gamma(T_j, T_{j+1}) = \rho\sqrt{S_B(T_j, T_{j+1})Q_B(T_j, T_{j+1})S_A(T_j, T_{j+1})Q_A(T_j, T_{j+1})} \tag{25e}$$

Proof: See the Appendix.

The individual payoffs under Proposition 8 are coupled and cannot be evaluated separately. The present value takes into account the results of all future decisions. The valuation proceeds via backward induction.

Proposition 8 has a general form that applies in a particular situation where we assume that parties A and B do not default simultaneously and have independent default risks, i.e. $\rho = 0$ and $\varphi_{AB} = 0$.

**Corollary 8.1:** *If parties A and B do not default simultaneously and have independent default risks, the bilateral risky value of the single payment derivative under the DTM is given by*

$$V^D(t) = \sum_{i=1}^{m} E\left[\left(\prod_{j=0}^{i-1} \bar{Y}(T_j, T_{j+1})\right) X_i \bigg| \mathcal{F}_t\right] \quad (26)$$

where $\bar{Y}(T_j, T_{j+1})$ is defined in (22).

The proof of this corollary is easily obtained according to Proposition 8 by setting $\rho = 0$ and $\varphi_{AB} = 0$.

## 4. A Practical Framework For Bilateral Risky Valuation and CVA

The risky valuation theory described above can be applied to any defaultable derivatives. In this section, we develop a practical framework to demonstrate how to perform the risky valuation and how to calculate the bilateral CVA at a portfolio level. The framework incorporates netting and margin agreements and captures right/wrong way risk. We use the DTM as an example. Switching to the CTM is quite straightforward, but requires more granular simulation time steps, e.g. daily.

Two parties are denoted as A and B. All calculations are from the perspective of party A. Let the valuation date be *t*. The risky valuation and CVA computation procedure consists of the following steps.

### 4.1. Risk-Neutral Monte Carlo Scenario Generation

One core element of a counterparty credit risk management system is the Monte Carlo scenario generation (market evolution). This must be able to run a large number of scenarios for each risk factor with flexibility over parameterization of processes and treatment of correlation between underlying factors. Credit exposure may be calculated under a real probability measure, while CVA may be conducted under a risk-neutral probability measure.

Due to the extensive computational intensity of pricing counterparty credit risk, there will inevitably be some compromise regarding limiting the number of market scenarios (paths) and the number of simulation dates (also called "time buckets" or "time nodes"). The time buckets are normally designed as fine-granularity at the short end and coarse-granularity at the far end. The details of scenario generation are beyond the scope of this paper.

### 4.2. Cash Flow Generation

For ease of illustration, we choose a vanilla interest rate swap as an example. For most banks, interest rate swaps account for more than half of their derivative business.

Assume that party *A* pays a fixed rate, while party *B* pays a floating-rate. We are considering that fixed rate payments and floating-rate payments occur at the same payment dates and with the same day-count conventions, and ignoring the swap funding spread. Though the generalization to different payment dates, day-count conventions and swap funding spreads is straightforward, we prefer to present a simplified version to ease the notation.

Assume that there are *M* time buckets $(T_0, T_1, ..., T_M)$ in each scenario, and *N* cash flows in the sample swap. Let us consider scenario *j* first.

For swaplet *i*, there are four important dates: the fixing date $t_{i,f}$, the starting date $t_{i,s}$, the ending date $t_{i,e}$ and the payment date $t_{i,p}$. In general, these dates are not coincidently at the simulation time buckets. The time relationship between swaplet *i* and the simulation time buckets is shown in Figure 1.

The cash flow generation consists of two procedures: cash flow determination and cash flow allocation.

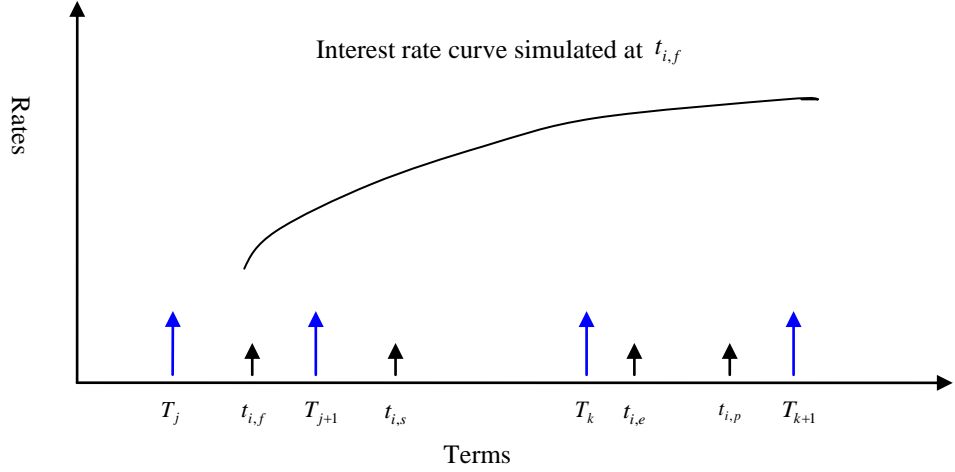

**Figure 1: Time Relationship between the Swaplet and the Simulation Time Buckets**

The floating leg of the interest rate swaplet is reset at the fixing date $t_{i,f}$ with the starting date $t_{i,s}$, the ending date $t_{i,e}$, and the payment date $t_{i,p}$. The simulation time buckets are $T_i, T_{i+1},...,T_{k+1}$. The simulated interest rate curve starts at $t_{i,f}$.

### 4.2.1. Cash Flow Determination

The cash flow of swaplet *i* is determined at the fixing date $t_{i,f}$, which is assumed to be between the simulation time buckets $T_j$ and $T_{j+1}$. First, we need to create an interest rate curve, observed at $t_{i,f}$, by interpolating the interest rate curves simulated at $T_j$ and $T_{j+1}$ via either Brownian Bridge or linear interpolation. Then, we can calculate the payoff of swaplet *i* at scenario *j* as

$$\chi_{j,i} = N\big(F(t_{i,f};t_{i,s},t_{i,e}) - R\big)\delta(t_{i,s},t_{i,e}) \tag{27}$$

where N denotes the notional, $F(t_{i,f};t_{i,s},t_{i,e})$ denotes the simply compounded forward rate reset at $t_{i,f}$ for the forward period ($t_{i,s}$, $t_{i,e}$), $\delta(t_{i,s},t_{i,e})$ denotes the accrual factor or day count fraction for the period ($t_{i,s}$, $t_{i,e}$), and *R* denotes the fixed rate.

### 4.2.2. Cash Flow Allocation

The cash flow amount calculated by (27) is paid on the payment date $t_{i,p}$. This value should be allocated into *the nearest previous time bucket* $T_k$ as:

$$\tilde{\chi}_{j,k,i} = \chi_{j,i} D(T_k, t_{i,p}) \tag{28}$$

where $D(T_k, t_{i,p})$ denotes the risk-free discount factor based on the interest rate curve simulated at $T_k$.

Cash flow generation for products without early-exercise provision is quite straightforward. For early-exercise products, one can use the approach proposed by Longstaff and Schwartz (2001) to obtain the optimal exercise boundaries, and then the payoffs.

### 4.3. Aggregation and Netting Agreements

After generating cash flows for each deal, we need to aggregate them at the counterparty portfolio level for each scenario and at each time bucket. The cash flows are aggregated by either netting or non-netting based on the netting agreements. A netting agreement is a provision that allows the offset of settlement payments and receipts on all contracts between two counterparties. Another important use of netting is close-out netting, which allows the offset of close-out values.

For netting, we add all cash flows together in the same scenario and at the same time bucket to recognize offsetting. The aggregated cash flow under netting at scenario *j* and time bucket *k* is given by

$$\tilde{\chi}_{j,k} = \sum_i \tilde{\chi}_{j,k,i} \tag{29a}$$

For non-netting, we divide cash flows into positive and negative groups and add them separately. In other words, offsetting is not recognized. The aggregated cash flows under non-netting at scenario *j* and time bucket *k* are given by

$$\tilde{\chi}_{j,k} = \begin{cases} \sum_l \tilde{\chi}_{j,k,l} & \text{if } \chi_{j,k,l} \geq 0 \\ \sum_m \tilde{\chi}_{j,k,m} & \text{if } \chi_{j,k,m} < 0 \end{cases} \tag{29b}$$

### 4.4. Margin (or Collateral) Agreements

Under a margin agreement, the collateral is called as soon as the counterparty exposure rises above the given collateral threshold $H$, or more precisely, above the threshold (*TH*) plus minimum transferable amount (*MTA*). This would result in a reduction of exposure by the collateral amount held $\Gamma$. Consequently, there would be no exposure above the threshold $H$ if there were no time lags between collateral calling, posting, liquidating, and closing out. However, these lags, which are actually the margin period of risk, do exist in practice. The collateral can depreciate or appreciate in value during this period. These lags expose the bank to additional exposure above the threshold, which is normally referred to as collateralized exposure. Clearly, the longer the margin period of risk, the larger the collateralized exposure.

Assume that the collateral margin period of risk is $\varsigma$. The collateral methodology consists of the following procedures:

First, for any time bucket $T_k$, we introduce an additional collateral time node $T_k - \varsigma$. Second, we compute the portfolio value $V_j^F(T_k - \varsigma)$ at scenario $j$ and collateral time node $(T_k - \varsigma)$. Then, we calculate the collateral required to reduce the exposure at $(T_k - \varsigma)$ as

$$\Gamma_j(T_k - \varsigma) = \begin{cases} V_j^F(T_k - \varsigma) - H_B, & \text{if } V_j^F(T_k - \varsigma) \geq H_B \\ 0, & \text{if } H_A < V_j^F(T_k - \varsigma) < H_B \\ V_j^F(T_k - \varsigma) - H_A & \text{if } V_j^F(T_k - \varsigma) \leq H_A \end{cases} \tag{30}$$

where $H_B = TH_B + MTA_B$ is the collateral threshold of party *B* and $H_A = -(TH_A + MTA_A)$ is the negative collateral threshold of party *A*.

Each bank has its own collateral simulation methodology that simulates the collateral value evolving from $\Gamma_j(T_k - \varsigma)$ to $\Gamma_j(T_k)$ over the margin period of risk $\varsigma$. The details of collateral simulation are beyond the scope of this paper. Assume that the collateral value $\Gamma_j(T_k)$ has already been calculated.

Next, we compute the change amount of the collaterals between $T_k$ and $T_{k-1}$ as

$$\Delta\Gamma_{j,k} := \Delta\Gamma_j(T_k) = \Gamma_j(T_k) - \Gamma_j(T_{k-1}) \tag{31}$$

Since our CVA methodology is based on cash flows, we model collateral as a reversing cash flow $-\Delta\Gamma_{j,k}$ at $T_k$. Finally, the total cash flow at $T_k$ is given by

$$\bar{\chi}_{j,k} = \tilde{\chi}_{j,k} - \Delta\Gamma_{j,k} \qquad (32)$$

**4.5. Wrong or Right Way Risk**

Wrong way risk occurs when exposure to a counterparty is adversely correlated with the credit quality of that counterparty, while right way risk occurs when exposure to a counterparty is positively correlated with the credit quality of that counterparty. For example, in wrong way risk, exposure tends to increase when counterparty credit quality worsens, while in right way risk, exposure tends to decrease when counterparty credit quality declines. Wrong/right way risk, as an additional source of risk, is rightly of concern to banks and regulators.

To capture wrong/right way risk, we need to correlate the credit quality (credit spreads or hazard rates) with other market risk factors, e.g. equities, commodities, etc., in the scenario generation.

**4.6. CVA Calculation**

After aggregating all cash flows via netting and margin agreements, one can price a portfolio in the same manner as pricing a single deal. We assume that the reader is familiar with the regression-based Monte Carlo valuation model proposed by Longstaff and Schwartz (2001) and thus do not repeat some well-known procedures for brevity.

4.6.1. Risk-Free Valuation

We first calculate the risk-free present value of a counterparty portfolio. The risk-free value at scenario $j$ is given by

$$V_j^F(t) = \sum_{k=1}^{m} \bar{\chi}_{j,k} D(t, T_k) \qquad (33a)$$

The final risk-free portfolio value is the average (expectation) of all scenarios given by

$$V^F(t) = E\big(V_j^F(t)\big) = E\bigg(\sum_{k=1}^{m} \bar{\chi}_{j,k} D(t, T_k)\bigg) \qquad (33b)$$

4.6.2. Risky Valuation

The risky valuation procedure is performed iteratively, starting at the last effective time bucket $T_m$, and then working backwards towards the present. We know the value of the portfolio at the final effective time bucket, which is equal to the last cash flow, i.e. $V_j^D(T_m) = \bar{\chi}_{j,m}$.

Based on the sign of $V_j^D(T_m)$ and Proposition 8, we can choose a proper risk-adjusted discount factor. The discounted cash flow at $T_{m-1}$ is given by

$$\Psi_{j,m-1} = Y_j(T_{m-1},T_m)V_j^D(T_m) = Y_j(T_{m-1},T_m)\bar{\chi}_{j,m} \tag{34a}$$

where

$$Y_j(T_{m-1},T_m) = D_j(T_{m-1},T_m)\left(1_{V_j^D(T_m)\geq 0}\, y_{j,B}(T_{m-1},T_m) + 1_{V_j^D(T_m)<0}\, y_{j,A}(T_{m-1},T_m)\right) \tag{34b}$$

where $y_{j,B}(T_{m-1},T_m)$ and $y_{j,A}(T_{m-1},T_m)$ are defined in (25).

Let us go to the penultimate effective time bucket $T_{m-1}$. The risk-adjusted discount factor has a switching type, depending on the sign of $V_j^D(T_{m-1}) + \bar{\chi}_{j,m-1}$, where $V_j^D(T_{m-1})$ is the value of the portfolio excluding the current cash flow $\bar{\chi}_{j,m-1}$ at scenario $j$ and time bucket $T_{m-1}$. Note that $V_j^D(T_{m-1})$ is not the discounted cash flow, but rather the expectation of the discounted cash flow conditional on the market states. We can use the well-known regression approach proposed by Longstaff and Schwartz (2001) to estimate $V_j^D(T_{m-1})$ from cross-sectional information in the simulation by using least squares. After estimating $V_j^D(T_{m-1})$, we can express the discounted cash flow at $T_{m-2}$ as

$$\Psi_{j,m-2} = Y_j(T_{m-2},T_{m-1})Y_j(T_{m-1},T_m)\bar{\chi}_{j,m} + Y_j(T_{m-2},T_{m-1})\bar{\chi}_{j,m-1} \tag{35a}$$

where

$$Y_j(T_{m-2},T_{m-1}) = D_j(T_{m-2},T_{m-1})\left(1_{(\bar{\chi}_{j,m-1}+V_j^D(T_{m-1}))\geq 0}\, y_{j,B}(T_{m-2},T_{m-1}) + 1_{\bar{\chi}_{j,m-1}+V_j^D(T_{m-1}))<0}\, y_{j,A}(T_{m-2},T_{m-1})\right) \tag{35b}$$

where $y_{j,B}(T_{m-2},T_{m-1})$ and $y_{j,A}(T_{m-2},T_{m-1})$ are defined in (25).

Assume that in the previous step $T_i$, we estimated the portfolio value $V_j^D(T_i)$. Now, we proceed to $T_{i-1}$. The discounting switch-type depends on the sign of $V_j^D(T_i) + \bar{\chi}_{j,i}$. The discounted cash flow at $T_{i-1}$ is given by

$$\Psi_{j,i-1} = \sum_{k=i}^{m} \left( \prod_{l=i-1}^{k-1} Y_j(T_l, T_{l+1}) \right) \bar{\chi}_{j,k} \tag{36}$$

We conduct the backward induction process, performed by iteratively rolling back a series of long jumps from the final effective time bucket $T_m$ across time nodes until we reach the valuation date. Then, the present value at scenario $j$ is

$$\Psi_{j,0} = \sum_{k=1}^{m} \left( \prod_{i=0}^{k-1} Y_j(T_i, T_{i+1}) \right) \bar{\chi}_{j,k} \tag{37a}$$

The final true/risky portfolio value is the average (expectation) of all scenarios, given by

$$V^D(t) = E(\Psi_{j,0}) = E\left[ \sum_{k=1}^{m} \left( \prod_{i=0}^{k-1} Y_j(T_i, T_{i+1}) \right) \bar{\chi}_{j,k} \right] \tag{37b}$$

CVA is by definition the difference between the risk-free portfolio value and the true (or risky or defaultable) portfolio value given by

$$CVA^B(t) = V^F(t) - V^D(t) = E\left\{ \sum_{k=1}^{m} \left[ \left( \prod_{i=0}^{k-1} D_j(T_i, T_{i+1}) \right) \bar{\chi}_{j,k} - \left( \prod_{i=0}^{k-1} Y_j(T_i, T_{i+1}) \right) \bar{\chi}_{j,k} \right] \right\} \tag{38}$$

## 5. Numerical Results

In this section, we present some numerical results for risky valuations and CVA calculations based on the theory and the practical framework described above.

### 5.1. Comparison Between the CTM and the DTM

The theoretical study of this article is conducted under both the CTM and the DTM. The CTM assumes that a default may occur at any time, while the DTM assumes that a default may only happen at discrete payment dates.

Here we use unilateral credit risk as an example. Assume that party $B$ has an 'A' rating and a constant default recovery rate of 70%. Party A is default-free. All calculations are from the perspective of party $A$.

Let us consider single payment derivatives first. We use a 6-month zero-coupon bond and a 1-year zero-coupon bond to analog single payment instruments. Under the CTM, the derivatives are continuously defaultable, whereas under the DTM, the derivatives may default only at 6 months or 1 year. The principals are a unit, and the hazard rates are bootstrapped from CDS spreads. The risk-free and risky values and the CVAs are calculated according to Corollary 1.2 and Corollary 2.1, shown in Table 2 and Table 3. The results demonstrate that the CVAs under both the CTM and the DTM are very close (relative difference < 0.3%).

Then, we use a 10-year semi-annual fixed rate coupon bond to analog a multiple payments instrument. Assume that the principal is a unit and that the annual coupon rate of the fixed-rate coupon bond is 4%. The risk-free and risky values and the CVAs are calculated according to Corollary 3.2 and Corollary 4.1, shown in Table 4. The results confirm that the DTM and the CTM produce very close results.

**Table 2. The Numerical Results of a 6-month Maturity Single Payment Derivative**

This table shows the present risk-free and unilateral risky values, and the unilateral CVAs of a 6-month maturity single payment derivative calculated under both the CTM and the DTM. The derivative has only one payoff at 6-month maturity and a unit principal. Difference of CVA = 1 – DTM-CVA / CTM-CVA.

| Risk-Free Value | CTM | | DTM | | Difference of CVA |
|---|---|---|---|---|---|
| | Risky Value | CVA | Risky Value | CVA | |
| 0.998168 | 0.997026 | 0.001142 | 0.997028 | 0.001140 | 0.1334% |

**Table 3. The Numerical Results of a 1-year Maturity Single Payment Derivative**

This table shows the present risk-free and unilateral risky values, and the unilateral CVAs of a 1-year maturity single payment derivative computed under both the CTM and the DTM. The derivative has only one payoff at 1-year maturity and a unit principal. Difference of CVA = 1 – DTM-CVA / CTM-CVA.

| Risk-Free Value | CTM | | DTM | | Difference of CVA |
| --- | --- | --- | --- | --- | --- |
| | Risky Value | CVA | Risky Value | CVA | |
| 0.995693 | 0.993422 | 0.002271 | 0.993428 | 0.002265 | 0.2658% |

**Table 4 The Numerical Results of a 10-Year Maturity Multiple Payments Derivative**

This table shows the present risk-free and unilateral risky values, and the unilateral CVAs of a 10-year maturity multiple payments derivative calculated under both the CTM and the DTM. The derivative has a unit principal and semi-annual payments with an annual coupon rate of 4%. Difference of CVA = 1 – DTM-CVA / CTM-CVA.

| Risk-Free Value | CTM | | DTM | | Difference of CVA |
| --- | --- | --- | --- | --- | --- |
| | Risky Value | CVA | Risky Value | CVA | |
| 1.122892 | 1.032344 | 0.090547 | 1.032901 | 0.089990 | 0.6153% |

### 5.2. Impact of Margin Agreements

In this section, we use a portfolio to study the impact of margin agreements. The portfolio consists of a number of derivatives on interest rate, equity, and foreign exchange. In the real world, it is very rare that a portfolio is positive all the time and in all scenarios. The number of simulation scenarios (or paths) is 20,000. The time buckets are set weekly. If the computational requirements exceed the system limit, one can reduce both the number of scenarios and the number of time buckets. The time buckets can be designed as fine-granularity at the short end, and coarse-granularity at the far end, e.g. daily, weekly, monthly and yearly, etc. The rationale is that calculations become less accurate due to accumulated errors from simulation discretization and inherited errors from the calibration of the underlying models, such as those due to a change in macro-economic climate. The collateral margin period of risk is assumed to be 14 days (2 weeks).

We use the CIR (Cox-Ingersoll-Ross) models for interest rate and hazard rate scenario generations, the modified GBM (Geometric Brownian Motion) models for equity and collateral evolutions, and the BK (Black Karasinski) models for foreign exchange dynamics. Table 5 illustrates that if party A has an infinite collateral threshold $H_A = \infty$ i.e. no collateral requirement on A, the bilateral CVA value increases, while the threshold $H_B$ increases. Table 6 shows that if party B has an infinite collateral threshold $H_B = \infty$, the bilateral CVA value actually decreases, while the threshold $H_A$ increases. This reflects the bilateral impact of the collaterals on the bilateral CVA. The impact is mixed in Table 7 when both parties have finite collateral thresholds.

**Table 5. The Impact of Collateral Threshold $H_B$ on the Bilateral CVA**

$H_B$ denotes the collateral threshold of party B and $H_A$ denotes the collateral threshold of party A. We set $H_A = \infty$ and change $H_B$ only.

| Collateral Threshold $H_B$ | 10.1 Mil | 15.1 Mil | 20.1 Mil | Infinite ($\infty$) |
|---|---|---|---|---|
| CVA | 19,550.91 | 20,528.65 | 21,368.44 | 22,059.30 |

**Table 6. The Impact of Collateral Threshold $H_A$ on the Bilateral CVA**

$H_B$ denotes the collateral threshold of party B and $H_A$ denotes the collateral threshold of party A. We set $H_B = \infty$ and change $H_A$ only.

| Collateral Threshold $H_A$ | 10.1 Mil | 15.1 Mil | 20.1 Mil | Infinite ($\infty$) |
|---|---|---|---|---|
| CVA | 28,283.64 | 25,608.92 | 23,979.11 | 22,059.30 |

**Table 7. The Impact of Both Collateral Thresholds on the Bilateral CVA**

$H_B$ denotes the collateral threshold of party B and $H_A$ denotes the collateral threshold of party A. Both of them are changed.

| Collateral Threshold $H_B$ | 10.1 Mil | 15.1 Mil | 20.1 Mil | Infinite ($\infty$) |
|---|---|---|---|---|
| Collateral Threshold $H_A$ | 10.1 Mil | 15.1 Mil | 20.1 Mil | Infinite ($\infty$) |
| CVA | 25,752.98 | 22,448.45 | 23,288.24 | 22,059.30 |

### 5.3. Impact of Wrong or Right Way Risk

We use an equity swap as an example. Assume the correlation between the underlying equity price and the credit quality (hazard rate) of party $B$ is $\rho$. The impact of the correlation on the CVA is show in Table 8. The results state that the CVA increases when the absolute value of the negative correlation increases.

**Table 8. Impact of Wrong Way Risk on the Bilateral CVA**

We use an equity swap as an example and assume that there is a negative correlation between the equity price and the credit quality of party $B$.

| Correlation $\rho$ | 0 | -50% | -100% |
|---|---|---|---|
| CVA | 165.15 | 205.95 | 236.99 |

## 6. Conclusion

This article presents a theory for pricing defaultable financial derivatives and their CVAs. First, we want to indicate that the market value of a defaultable derivative is actually a risky value rather than a risk-free value. In fact, in applying the upfront CVA, we already converted the market value of a defaultable derivative from the risk-free value to the risky value.

For completeness, our theoretical study covers various cases. We find that the valuation of defaultable derivatives and their CVAs, in most situations, has a backward recursive nature and requires a backward induction valuation. An intuitive explanation is that two counterparties implicitly sell each other an option to default when entering into a defaultable transaction. If we assume that a default may occur at any time, the default options are American style options. If we assume that a default may only happen on the payment dates, the default options are either European options or Bermudan options. Both Bermudan and American options require backward induction valuations.

Based on our theory, we propose a novel cash-flow-based practical framework for calculating the bilateral risky value and bilateral CVA at the counterparty portfolio level. This framework can easily incorporate various credit mitigation techniques, such as netting agreements and margin agreements, and can capture wrong/right way risk. Numerical results show that these credit mitigation techniques and wrong/right way risk have significant impacts on CVA.

## Appendix

**Proof of Proposition 1:** Under the unilateral credit risk assumption, only the default risk of one party appears to be relevant, i.e., we only consider the default risk when the derivative is in the money. We divide the time period $(t, T)$ into $n$ very small intervals ($\Delta t$) and use the approximation $\exp(y) \approx 1 + y$ for very small $y$. Assume that a default may occur only at the end of each small period. The survival and the default probabilities for the period $(t, t + \Delta t)$ are given by

$$\breve{S}_B(t) := S_B(t, t + \Delta t) = \exp(-h_B(t)\Delta t) \approx 1 - h_B(t)\Delta t \tag{A1a}$$

$$\breve{Q}_B(t) := Q_B(t, t + \Delta t) = 1 - \exp(-h_B(t)\Delta t) \approx h_B(t)\Delta t \tag{A1b}$$

At time $t + \Delta t$ the derivative either defaults or survives. The risky value of the derivative at $t$ is given by

$$\begin{aligned} V^D(t) &= E\left\{\exp(-r(t)\Delta t)\left[1_{V^D(t+\Delta t)\geq 0}\left(\breve{S}_B(t) + \varphi_B(t)\breve{Q}_B(t)\right) + 1_{V^D(t+\Delta t)<0}\right]V^D(t+\Delta t)\big|\mathcal{F}_t\right\} \\ &\approx E\left\{\exp(-r(t)\Delta t)\left[1_{V^D(t+\Delta t)\geq 0}\left(1 - h_B(t)\Delta t + \varphi_B(t)h_B(t)\Delta t\right) + 1_{V^D(t+\Delta t)<0}\right]V^D(t+\Delta t)\big|\mathcal{F}_t\right\} \\ &\approx E\left\{\exp(-c_B(t)\Delta t)V^D(t+\Delta t)\big|\mathcal{F}_t\right\} \end{aligned} \tag{A2a}$$

where $1_Y$ is an indicator function that is equal to one if Y is true and zero otherwise; and

$$1_{V^D(t+\Delta t)\geq 0} + 1_{V^D(t+\Delta t)<0} = 1 \tag{A2b}$$

$$c_B(t) = r(t) + 1_{V^D(t+\Delta t)\geq 0}h_B(t)(1 - \varphi_B(t)) \tag{A2c}$$

Similarly, we have

$$V^D(t+\Delta t) = E\{\exp(-c_B(t+\Delta t)\Delta t)V^D(t+2\Delta t)|\mathcal{F}_{t+\Delta t}\} \tag{A3}$$

Note that $\exp(-c_B(t)\Delta t)$ is $\mathcal{F}_{t+\Delta t}$-measurable. By definition, an $\mathcal{F}_{t+\Delta t}$-measurable random variable is a random variable whose value is known at time $t+\Delta t$. Based on the *taking out what is known* and *tower* properties of conditional expectation, we have

$$\begin{aligned}V^D(t) &= E\{\exp(-c_B(t)\Delta t)V^D(t+\Delta t)|\mathcal{F}_t\} \\ &= E\{\exp(-c_B(t)\Delta t)E[\exp(-c_B(t+\Delta t)\Delta t)V^D(t+2\Delta t)|\mathcal{F}_{t+\Delta t}]|\mathcal{F}_t\} \\ &= E\{E[\exp(-\sum_{i=0}^{1}c_B(t+i\Delta t)\Delta t))V^D(t+2\Delta t)|\mathcal{F}_{t+\Delta t}]|\mathcal{F}_t\} \\ &= E\{\exp(-\sum_{i=0}^{1}c_B(t+i\Delta t)\Delta t))V^D(t+2\Delta t)|\mathcal{F}_t\}\end{aligned} \tag{A4}$$

By recursively deriving from $t$ forward over $T$, where $V^D(T) = X_T$, the price can be expressed as

$$V^D(t) = E\{\exp[-\sum_{i=0}^{n-1}c_B(t+i\Delta t)\Delta t]X_T|\mathcal{F}_t\} = E\{C_B(t,T)X_T|\mathcal{F}_t\} \tag{A5a}$$

where

$$c_B(t+i\Delta t) = r(t+i\Delta t) + 1_{V^D(t+(i+1)\Delta t)\geq 0}h_B(t+i\Delta t)(1-\varphi_B(t+i\Delta t)) \tag{A5b}$$

**Proof of Proposition 2:** Under the unilateral credit risk assumption, we only consider the default risk when the derivative is in the money. Assume that a default may only occur on the payment date. Therefore, the risky value of the derivative at $t$ is given by

$$\begin{aligned}V^D(t) &= E\{D(t,T)[1_{X_T\geq 0}(S_B(t,T)+\varphi_B(T)Q_B(t,T))+1_{X_T<0}]X_T|\mathcal{F}_t\} \\ &= E\{D(t,T)[1-1_{X_T\geq 0}(1-\varphi_B(T))Q_B(t,T)]X_T|\mathcal{F}_t\} = E[G_B(t,T)X_T|\mathcal{F}_t]\end{aligned} \tag{A6a}$$

where

$$1_{X_T\geq 0} + 1_{X_T<0} = 1 \tag{A6b}$$

$$S_B(t,T) = 1 - Q_B(t,T) \tag{A6c}$$

$$G_B(t,T) = D(t,T)[1 - 1_{X_T\geq 0}Q_B(t,T)(1-\varphi_B(T))] \tag{A6d}$$

**Proof of Proposition 3:** We divide any payment date period ($T_{i-1}$, $T_i$) into $n_i$ very small time intervals ($\Delta t$). On the first cash flow payment date $T_1$, let $V^D(T_1)$ denote the value of the risky derivative excluding the current cash flow $X_1$. Both $V^D(T_1)$ and $X_1$ could be positive or negative. According to Proposition 1, the risky value of the derivative at $t$ is given by

$$V^D(t) = E[C_B(t,T_1)(X_1 + V^D(T_1))|\mathcal{F}_t] \tag{A7}$$

where $C_B(t,T_1)$ is defined in (A5)

Similarly, we have

$$V^D(T_1) = E[C_B(T_1,T_2)(X_2 + V^D(T_2))|\mathcal{F}_{T_1}] \tag{A8}$$

Note that $C_B(t,T_1)$ is $\mathcal{F}_{T_1}$-measurable. According to the *taking out what is known* and *tower* properties of conditional expectation, we have

$$\begin{aligned} V^D(t) &= E[C_B(t,T_1)(X_1 + V^D(T_1))|\mathcal{F}_t] = E[C_B(t,T_1)X_1|\mathcal{F}_t] \\ &+ E\{C_B(t,T_1)[E(C_B(T_1,T_2)X_2|\mathcal{F}_{T_1}) + E(C_B(T_1,T_2)V^D(T_2)|\mathcal{F}_{T_1})]|\mathcal{F}_t\} \\ &= \sum_{i=1}^{2} E\{C_B(t,T_i)X_i|\mathcal{F}_t\} + E[C_B(t,T_2)V^D(T_2)|\mathcal{F}_t] \end{aligned} \tag{A9a}$$

where

$$C_B(t,T_2) = C_B(t,T_1)C_B(T_1,T_2) = \exp\left[-\sum_{j=0}^{n_1+n_2-1} c_B(t+j\Delta t)\Delta t\right] \tag{A9b}$$

By recursively deriving from $T_2$ forward over $T_m$, where $V^D(T_m) = X_m$, we have

$$V^D(t) = \sum_{i=1}^{m} E[(C_B(t,T_i))X_i|\mathcal{F}_t] \tag{A10}$$

**Proof of Proposition 4:** Let $t = T_0$. Assume that a default may only occur on the payment dates. On the first payment date $T_1$, let $V^D(T_1)$ denote the risky value of the derivative excluding the current cash flow $X_1$. According to Proposition 2, the risky value of the derivative at $t$ is given by

$$V^D(t) = E[G_B(T_0,T_1)(X_1 + V^D(T_1))|\mathcal{F}_t] \tag{A11a}$$

where

$$G_B(T_0,T_1) = D(T_0,T)\left[1 - 1_{(X_1+V^D(T_1))\geq 0} Q_B(T_0,T_1)(1-\varphi_B(T_1))\right] \tag{A11b}$$

Similarly, we have

$$V^D(T_1) = E\left[G_B(T_1,T_2)(X_2 + V^D(T_2))\big|\mathcal{F}_{T_1}\right] \quad (A12)$$

Note that $G_B(T_0,T_1)$ is $\mathcal{F}_{T_1}$-measurable. According to the *taking out what is known* and *tower* properties of conditional expectation, we have

$$\begin{aligned}V^D(t) &= E\left[G_B(T_0,T_1)(X_1 + V^D(T_1))\big|\mathcal{F}_t\right] = E\left[G_B(T_0,T_1)X_1\big|\mathcal{F}_t\right] \\ &\quad + E\left\{G_B(T_0,T_1)\left[E\left(G_B(T_1,T_2)X_2\big|\mathcal{F}_{T_1}\right) + E\left(G_B(T_1,T_2)V^D(T_2)\big|\mathcal{F}_{T_1}\right)\right]\big|\mathcal{F}_t\right\} \\ &= \sum_{i=1}^{2} E\left[\left(\prod_{j=0}^{i-1} G_B(T_j,T_{j+1})\right)X_i\big|\mathcal{F}_t\right] + E\left[\left(\prod_{j=0}^{1} G_B(T_j,T_{j+1})\right)V^D(T_2)\big|\mathcal{F}_t\right]\end{aligned} \quad (A13)$$

By recursively deriving from $T_2$ forward over $T_m$, where $V^D(T_m) = X_m$, we have

$$V^D(t) = \sum_{i=1}^{m} E\left[\left(\prod_{j=0}^{i-1} G_B(T_j,T_{j+1})\right)X_i\big|\mathcal{F}_t\right] \quad (A14)$$

**Proof of Proposition 5:** We divide the time period (*t, T*) into *n* very small intervals ($\Delta t$) and use the approximation $\exp(y) \approx 1 + y$ for very small *y*. The survival and the default probabilities of party *j (j=A, B)* for the period (*t*, $t+\Delta t$) are given by

$$\breve{S}_j(t) := S_j(t, t+\Delta t) = \exp(-h_j(t)\Delta t) \approx 1 - h_j(t)\Delta t$$

$$\breve{Q}_j(t) := Q_j(t, t+\Delta t) = 1 - \exp(-h_j(t)\Delta t) \approx h_j(t)\Delta t$$

At time $t+\Delta t$, there are four possible states: 1) both *A* and *B* survive, 2) *A* defaults but *B* survives, 3) *A* survives but *B* defaults, and 4) both *A* and *B* default. The joint distribution of *A* and *B* is shown in Table 1. Depending on whether the market value of the derivative is an asset or a liability at $t+\Delta t$, we have

$$
\begin{aligned}
V^D(t) &= E\bigg\{\exp(-r(t)\Delta t)\bigg[1_{V^D(t+\Delta t)\geq 0}\Big\langle \big(\overset{)}{S}_B(t)\overset{)}{S}_A(t)+\gamma(t)\big)+\varphi_B(t)\big(\overset{)}{Q}_B(t)\overset{)}{S}_A(t)-\gamma(t)\big) \\
&\quad +\overline{\varphi}_B(t)\big(\overset{)}{S}_B(t)\overset{)}{Q}_A(t)-\gamma(t)\big)+\varphi_{AB}(t)\big(\overset{)}{Q}_B(t)\overset{)}{Q}_A(t)+\gamma(t)\big)\Big\rangle V^D(t+\Delta t)\Big|\mathcal{F}_t \\
&\quad +1_{V^D(t+\Delta t)<0}\Big\langle \big(\overset{)}{S}_B(t)\overset{)}{S}_A(t)+\gamma(t)\big)+\overline{\varphi}_A(t)\big(\overset{)}{Q}_B(t)\overset{)}{S}_A(t)-\gamma(t)\big) \\
&\quad +\varphi_A(t)\big(\overset{)}{S}_B(t)\overset{)}{Q}_A(t)-\gamma(t)\big)+\varphi_{AB}(t)\big(\overset{)}{Q}_B(t)\overset{)}{Q}_A(t)+\gamma(t)\big)\Big\rangle V^D(t+\Delta t)\Big|\mathcal{F}_t\bigg]\bigg\} \\
&\approx E\bigg\{\exp(-r(t)\Delta t)\bigg[1_{V^D(t+\Delta t)\geq 0}\Big\langle 1-h_A(t)(1-\overline{\varphi}_B(t))\Delta t - h_B(t)(1-\varphi_B(t))\Delta t \\
&\quad +\big(1-\varphi_B(t)-\overline{\varphi}_B(t)+\varphi_{AB}(t)\big)\big(\gamma(t)+h_B(t)h_A(t)\Delta t^2\big)\Big\rangle V^D(t+\Delta t)\Big|\mathcal{F}_t \\
&\quad +1_{V^D(t+\Delta t)<0}\Big\langle 1-h_A(t)(1-\varphi_A(t))\Delta t - h_B(t)(1-\overline{\varphi}_A(t))\Delta t \\
&\quad +\big(1-\varphi_A(t)-\overline{\varphi}_A(t)+\varphi_{AB}(t)\big)\big(\gamma(t)+h_B(t)h_A(t)\Delta t^2\big)\Big\rangle V^D(t+\Delta t)\Big|\mathcal{F}_t\bigg]\bigg\} \\
&\approx E\big\{\exp(-o(t)\Delta t)V^D(t+\Delta t)\big|\mathcal{F}_t\big\}
\end{aligned}
$$
(A15a)

where

$$o(t) = r(t) + 1_{V^D(t+\Delta t)\geq 0}\, p_B(t) + 1_{V^D(t+\Delta t)<0}\, p_A(t) \tag{A15b}$$

$$
\begin{aligned}
p_B(t) &= (1-\varphi_B(t))h_B(t) + (1-\overline{\varphi}_B(t))h_A(t) - (1-\varphi_B(t)-\overline{\varphi}_B(t)+\varphi_{AB}(t)) \\
&\quad \times \Big(\rho\sqrt{h_A(t)h_B(t)(1-h_B(t)\Delta t)(1-h_A(t)\Delta t)} + h_B(t)h_A(t)\Delta t\Big)
\end{aligned}
$$
(A15c)

$$
\begin{aligned}
p_A(t) &= (1-\varphi_A(t))h_A(t) + (1-\overline{\varphi}_A(t))h_B(t) - (1-\varphi_B(t)-\overline{\varphi}_B(t)+\varphi_{AB}(t)) \\
&\quad \times \Big(\rho\sqrt{h_A(t)h_B(t)(1-h_B(t)\Delta t)(1-h_A(t)\Delta t)} + h_B(t)h_A(t)\Delta t\Big)
\end{aligned}
$$
(A15d)

$$\gamma(t) = \rho\sqrt{h_A(t)h_B(t)(1-h_B(t)\Delta t)(1-h_A(t)\Delta t)}\,\Delta t^2 \tag{A15e}$$

where $\varphi_j$ ($j=A, B$) represents the default recovery rate of party $j$; $h_j$ represents the hazard rate of party $j$; $\overline{\varphi}_j$ represents the non-default recovery rate of party $j$. $\overline{\varphi}_j = 0$ represents the one-way settlement rule, while $\overline{\varphi}_j = 1$ represents two-way settlement rule. $\rho$ denotes the default correlation of $A$ and $B$. $\varphi_{AB}$ represents the joint recovery rate when $A$ and $B$ default simultaneously.

Similarly, we have

$$V^D(t+\Delta t) = E\big\{\exp(-o(t+\Delta t)\Delta t)V^D(t+2\Delta t)\big|\mathcal{F}_{t+\Delta t}\big\} \tag{A16}$$

Note that $\exp(-o(t)\Delta t)$ is $\mathcal{F}_{t+\Delta t}$-measurable. Based on the *taking out what is known* and *tower* properties of conditional expectation, we have

$$\begin{aligned}
V^D(t) &= E\{\exp(-o(t)\Delta t)V^D(t+\Delta t)|\mathcal{F}_t\} \\
&= E\{\exp(-o(t)\Delta t)E[\exp(-o(t+\Delta t)\Delta t)V^D(t+2\Delta t)|\mathcal{F}_{t+\Delta t}]|\mathcal{F}_t\} \\
&= E\{\exp(-\sum_{i=0}^{1}o(t+i\Delta t)\Delta t)V^D(t+2\Delta t)|\mathcal{F}_t\}
\end{aligned} \tag{A17}$$

By recursively deriving from $t$ forward over $T$, where $V^D(T) = X_T$, the price can be expressed as:

$$V^D(t) = E\{\exp(-\sum_{i=0}^{n-1}o(t+i\Delta t)\Delta t)X_T|\mathcal{F}_t\} = E\{O(t,T)X_T|\mathcal{F}_t\} \tag{A18}$$

**Proof of Proposition 6:** We assume that a default may only occur on the payment date. At time $T$, there are four possible states: 1) both $A$ and $B$ survive, 2) $A$ defaults but $B$ survives, 3) $A$ survives but $B$ defaults, and 4) both $A$ and $B$ default. The joint distribution of $A$ and $B$ is shown in Table 1. Depending on whether the payoff is in the money or out of the money at $T$, we have

$$\begin{aligned}
V^D(t) &= E\{D(t,T)[1_{X_T \geq 0}\langle (S_B(t,T)S_A(t,T)+\gamma(T)) + \varphi_B(T)(Q_B(t,T)S_A(t,T)-\gamma(T)) \\
&\quad + \overline{\varphi}_B(T)(S_B(t,T)Q_A(t,T)-\gamma(T)) + \varphi_{AB}(T)(Q_B(t,T)Q_A(t,T)+\gamma(T))\rangle X_T|\mathcal{F}_t \\
&\quad + 1_{X_T<0}\langle (S_B(t,T)S_A(t,T)+\gamma(T)) + \overline{\varphi}_A(T)(Q_B(t,T)S_A(t,T)-\gamma(T)) \\
&\quad + \varphi_A(T)(S_B(t,T)Q_A(t,T)-\gamma(T)) + \varphi_{AB}(T)(Q_B(t,T)Q_A(t,T)+\gamma(T))\rangle X_T|\mathcal{F}_t]\} \\
&= E(Y(t,T)X_T|\mathcal{F}_t)
\end{aligned} \tag{A19a}$$

where

$$Y(t,T) = D(t,T)(1_{X_T \geq 0}y_B(t,T) + 1_{X_T<0}y_A(t,T)) \tag{A19b}$$

$$\begin{aligned}
y_B(t,T) &= S_B(t,T)S_A(t,T) + \varphi_B(T)Q_B(t,T)S_A(t,T) + \overline{\varphi}_B(T)S_B(t,T)Q_A(t,T) \\
&\quad + \varphi_{AB}(T)Q_B(t,T)Q_A(t,T) + \gamma(t,T)(1-\varphi_B(T)-\overline{\varphi}_B(T)+\varphi_{AB}(T))
\end{aligned} \tag{A19c}$$

$$\begin{aligned}
y_A(t,T) &= S_B(t,T)S_A(t,T) + \varphi_A(T)Q_A(t,T)S_B(t,T) + \overline{\varphi}_A(T)S_A(t,T)Q_B(t,T) \\
&\quad + \varphi_{AB}(T)Q_B(t,T)Q_A(t,T) + \gamma(t,T)(1-\varphi_B(T)-\overline{\varphi}_B(T)+\varphi_{AB}(T))
\end{aligned} \tag{A19d}$$

$$\gamma(t,T) = \rho\sqrt{S_B(t,T)Q_B(t,T)S_A(t,T)Q_A(t,T)} \tag{A19e}$$

**Proof of Proposition 7.** On the first cash flow payment date $T_1$, let $V^D(T_1)$ denote the risky value of the derivative excluding the current cash flow $X_1$. According to Proposition 5, we have

$$V^D(t) = E[O(t,T_1)(X_1 + V^D(T_1))|\mathcal{F}_t] \tag{A20a}$$

Where $O(t,T_1)$ is defined in (A18) and (A15).

Similarly, we have

$$V^D(T_1) = E\left[O(T_1,T_2)(X_2 + V^D(T_2))\big|\mathcal{F}_{T_1}\right] \quad (A21)$$

Note that $O(t,T_1)$ is $\mathcal{F}_{T_1}$-measurable. According to the *taking out what is known* and *tower* properties of conditional expectation, we have

$$\begin{aligned}V^D(t) &= E\left[O(t,T_1)(X_1 + V^D(T_1))\big|\mathcal{F}_t\right] = E\left[O(t,T_1)X_1\big|\mathcal{F}_t\right] \\ &\quad + E\left\{O(t,T_1)\left[E(O(T_1,T_2)X_2|\mathcal{F}_{T_1}) + E(O(T_1,T_2)V^D(T_2)|\mathcal{F}_{T_1})\right]\big|\mathcal{F}_t\right\} \\ &= \sum_{i=1}^{2} E\left[O(t,T_i)X_i\big|\mathcal{F}_t\right] + E\left[O(t,T_2)V^D(T_2)\big|\mathcal{F}_t\right]\end{aligned} \quad (A22a)$$

where

$$O(t,T_2) = O(t,T_1)O(T_1,T_2) \quad (A22b)$$

By recursively deriving from $T_2$ forward over $T_m$, where $V^D(T_m) = X_m$, we have

$$V^D(t) = \sum_{i=1}^{m} E\left[O(t,T_i)X_i\big|\mathcal{F}_t\right] \quad (A23)$$

**Proof of Proposition 8.** We assume that a default may only occur on the payment dates. Let $t = T_0$. On the first cash flow payment date $T_1$, let $V^D(T_1)$ denote the risky value of the derivative excluding the current cash flow $X_1$. According to Proposition 6, we have

$$V^D(t) = E\left[Y(T_0,T_1)(X_1 + V^D(T_1))\big|\mathcal{F}_t\right] \quad (A24a)$$

where

$$Y(T_0,T_1) = D(T_0,T_1)\left(1_{(X_1+V^D(T_1))\geq 0}\, y_B(T_0,T_1) + 1_{(X_1+V^D(T_1))<0}\, y_A(T_0,T_1)\right) \quad (A24b)$$

where $y_B(T_0,T_1)$ and $y_A(T_0,T_1)$ are defined in (A19).

Similarly, we have

$$V^D(T_1) = E\left[Y(T_1,T_2)(X_2 + V^D(T_2))\big|\mathcal{F}_{T_1}\right] \quad (A25)$$

Note that $Y(T_0,T_1)$ is $\mathcal{F}_{T_1}$-measurable. According to *taking out what is known* and *tower* properties of conditional expectation, we have

$$V^D(t) = E[Y(T_0,T_1)(X_1 + V^D(T_1))|\mathcal{F}_t] = E[Y(T_0,T_1)X_1|\mathcal{F}_t]$$
$$+ E\{Y(T_0,T_1)[E(Y(T_1,T_2)X_2|\mathcal{F}_{T_1}) + E(Y(T_1,T_2)V^D(T_2)|\mathcal{F}_{T_1})]|\mathcal{F}_t\} \quad (A26)$$
$$= E[Y(T_0,T_1)X_1|\mathcal{F}_t] + E[(\prod_{j=0}^{1}Y(T_j,T_{j+1}))X_2|\mathcal{F}_t] + E[(\prod_{j=0}^{1}Y(T_j,T_{j+1}))V^D(T_2)|\mathcal{F}_t]$$

By recursively deriving from $T_2$ forward over $T_m$, where $V^D(T_m) = X_m$, we have

$$V^D(t) = \sum_{i=1}^{m} E[(\prod_{j=0}^{i-1}Y(T_j,T_{j+1}))X_i|\mathcal{F}_t] \quad (A27)$$

Merton, Robert C., 1974, On the pricing of corporate debt: the risk structure of interest rates, Journal of Finance, 29, 449-470.

Pykhtin, Michael, and Steven Zhu, 2007, A guide to modeling counterparty credit risk, GARP Risk Review, July/August, 16-22.

Xiao, Tim, 2015, An accurate solution for credit value adjustment (CVA) and wrong way risk, Journal of Fixed Income, Summer 2015, 25(1), 84-95.39